\documentclass[12pt,preprint]{aastex}
\usepackage{psfig,graphicx}

\newcommand{\etal}{et al.}

\begin{document}




\title{Discovery of Massive Evolved Galaxies at $z > 3$ in the Hubble Ultra Deep Field\altaffilmark{1}}
\altaffiltext{1}{Based on observations with the NASA/ESA Hubble Space
Telescope, obtained at the Space Telescope Science Institute, which is operated
by the Association of Universities for Research in Astronomy, Inc., under NASA
contract NAS5--26555.}

\author{HSIAO-WEN CHEN\altaffilmark{2,3} and RONALD O. MARZKE\altaffilmark{4}}
\altaffiltext{2}{Center for Space Research, Massachusetts Institute of
Technology, Cambridge, MA 02139-4307, U.S.A. {\tt hchen@space.mit.edu}}
\altaffiltext{3}{Hubble Fellow}

\altaffiltext{4}{Department of Physics and Astronomy, San Francisco State University, San Francisco, CA 94132-4163, U.S.A. {\tt marzke@stars.sfsu.edu}}

\begin{abstract}

We have identified six early-type galaxies at $z>2.8$ in the central 5.76 
arcmin$^2$ Hubble Ultra Deep Field NICMOS region based on a pronounced 
broad-band discontinuity between the NICMOS F110W and F160W bandpasses.  These 
galaxies have red observed optical and near-infrared colors ($AB_{\rm F775W} -
AB_{\rm F160W} \ge 2$) that cannot be explained entirely by dust reddening 
(indicating advanced age), and their luminosities at rest-frame optical 
wavelengths suggest a substantial stellar mass.  One of the galaxies is 
detected in X-ray and is likely to have an active nuclear region, while the 
rest are estimated to be {\em at least} 1 Gyr old at $z\approx 3$ and contain 
total stellar mass of $0.4 - 9.1 \times 10^{10}\,h^{-2}\,{\rm M_\odot}$.  We 
calculate a cumulative comoving stellar mass density of $\rho_{\rm star}=0.7 -
1.2\times 10^{7}\,h\,{\rm M_\odot}\,{\rm Mpc}^{-3}$ for massive evolved 
galaxies of $M > 10^{9}\,h^{-2}\,{\rm M_\odot}$ at $z>2.5$.  Comparison of the 
stellar mass density confined in galaxies of different types shows that massive
evolved galaxies contribute $\approx 16 - 28$\% of total stellar mass density 
even at these early epochs.  Finally, an analysis of their morphology observed 
in the ACS and NICMOS images shows that the active galaxy has compact 
structure, while the rest are less concentrated.  The presence of massive 
evolved galaxies at $z\sim 3$, when the universe was only 2 Gyr old, suggests 
that early star formation may have been particularly efficient in massive 
halos.

\end{abstract}


\section{INTRODUCTION}

The Hubble Ultra Deep Field (HUDF; Beckwith \etal\ 2004) imaging observations 
have recorded by far the deepest images of the distant universe using the 
Advanced Camera for Surveys (ACS).  These observations cover $3 \times 3$ 
arcmin$^2$ of sky area and reach $10\,\sigma$ detection limits of $AB=29.3, 
30.0, 29.7$, and 28.7 in the F435W, F606W, F775W, and F850LP bandpasses, 
respectively.  Near-infrared images of comparable depths in the inner $2.4
\times 2.4$ arcmin$^2$ region of the HUDF have also been obtained using the 
NICMOS camera (Thompson \etal\ 2004).  The NICMOS images reach $10\,\sigma$ 
detection limits of $AB=28.3$ in F110W and $AB=27.8$ in F160W, more than two 
magnitudes deeper than the deepest near-infrared images from the ground 
(e.g.\ Labb\'e \etal\ 2003).

A particularly exciting development is the opportunity to identify evolved
galaxies at epochs earlier than $z=2$, where the presence of massive galaxies
becomes difficult to explain as the result of time-consuming hierarchical
galaxy assembly (e.g.\ Baugh \etal\ 2002; Somerville \etal\ 2004).  Over the 
past two years, galaxy samples selected in the near-infrared have shown 
that evolved galaxies are not rare at $z>1$ (e.g.\ Chen \etal\ 2003; Pozzetti 
\etal\ 2003) and that at least 40\% of stellar mass density found locally was 
in place by $z=1$ (Fontana \etal\ 2003; Rudnick \etal\ 2003; Glazebrook \etal\ 
2004).  Together these results indicate that much of the assembly of stellar 
mass occurs at epochs much earlier than $z=1$.   

Detection of massive galaxies beyond $z = 2$ pushes back the epoch of 
early-type galaxy formation even further and thus offers stronger constraints 
on models of galaxy formation.  Deep ground-based imaging in the near-infrared 
has recently revealed a significant number of star-forming galaxies at 
$z=2.2-2.4$ with total stellar mass comparable to local early-type galaxies 
(Daddi \etal\ 2004; van Dokkum \etal\ 2004).  Here we report the detection of 
six massive, evolved galaxies identified at $\langle z\rangle=3.2$ in the 
HUDF/NICMOS images using photometric redshift techniques.  Our analysis of the 
photometric properties of these red galaxies indicates that they are evolved 
stellar populations and that some massive galaxies are already largely 
assembled by $z=3$.

We adopt a $\Lambda$ cosmology, $\Omega_{\rm M}=0.3$ and $\Omega_\Lambda = 
0.7$, with a dimensionless Hubble constant $h = H_0/(100 \ {\rm km} \ {\rm 
s}^{-1}\ {\rm Mpc}^{-1})$ throughout the paper.

\section{A CATALOG OF NEAR-INFRARED SELECTED HUDF GALAXIES}

We have performed an independent analysis of the drizzle-stacked optical and 
near-infrared images provided by the HUDF team and the NICMOS treasury team.
The object detection algorithm was similar to those described in Chen \etal\ 
(2002).  In summary, we first performed object detection in individual images 
using SExtractor (Bertin \& Arnouts 1996) and set the detection parameters such
that no detections were found in the negative images.  Catalogs of individual 
bandpasses were then combined to form a final catalog with flags indicating the
presence/absence of flux in respective bandpasses.  We identified 1833 objects 
of $AB({\rm F160W})\le 28.5$ over the central 5.76 arcmin$^2$ sky area.  

We have also measured the redshifts and redshift likelihood functions for all 
the objects using a photometric redshift analysis outlined in Chen \etal\ 
(2003).  The algorithm compares the observed SED, established from photometric 
measurements in the ACS F435W, F606W, F775W, F850LP, and NICMOS F110W and F160W
bandpasses, with a grid of model templates calculated at different redshifts.
We have considered a suite of galaxy templates that include E/S0, Sab, Scd,
Irr, two starbursts and a QSO templates, as well as a suite of stellar 
templates that range from early O stars through late-type T dwarfs.  
Photometric redshift techniques are particularly successful when either the 
intrinsic continuum absorption at the Lyman or 4000 \AA\ breaks or the external
continuum absorption due to the intervening Ly$\alpha$ forest are observed in 
the broad-band SEDs (e.g.\ Fern\'{a}ndez-Soto \etal\ 2001; Abraham \etal\ 
2004).  These broad features are easy to identify and are only mildly sensitive
to differences in the gas and dust content of the galaxies.

Figure 1 presents the observed F775W and F160W color ($I-H$) versus photometric
redshift for all galaxies in the HUDF observed with NICMOS.
While 
it is not surprising to find evolved galaxies with $AB_{\rm F775W} - AB_{\rm 
F160W} > 2$ at $z>1$ in the HUDF images, it is intriguing to see that nine such
galaxies lie at $z > 2.5$.  These objects exhibit a pronounced discontinuity 
between the F110W and F160W bandpasses that is matched in each case to the 
rest-frame 4000-\AA\ break in either an E/S0 or Sab galaxy template.


\section{PROPERTIES OF MASSIVE, OLD GALAXIES AT $z>2.8$}

Galaxies dominated by evolved stars are least affected by the details of recent
star formation and therefore allow the most reliable measurements of 
accumulated stellar mass.  Consequently, these evolved galaxies are 
particularly useful in discriminating between galaxy formation scenarios.  For 
hierarchical models (e.g.\ Kauffmann \& Charlot 1998), evolution in the number 
density of massive galaxies constrains the merging sequence, while the age of 
the underlying stellar population constrains the epoch when the first stars 
formed.

Our photometric redshift analysis identified nine early-type galaxies 
at $z>2.8$ by associating the pronounced broad-band discontinuity between 
the NICMOS F110W and F160W bandpasses with the rest-frame 4000-\AA\ break. 
This identification is the basis for our further analysis of the stellar 
populations and thus merits some scrutiny.  One concern is that strong 
reddening in intermediate-redshift galaxies could mimic the appearance of a 
high-redshift 4000-\AA\ break.  To examine whether the observed SEDs are the 
result of dust reddening, we repeated the photometric redshift analysis 
including additional dusty templates.  The dusty templates were generated using
the SB1 template (with intrinsic color excess $E(B-V)\le 0.1$) from Kinney 
\etal\ (1996), reddened according to the Calzetti extinction law (Calzetti, 
Kinney, \& Storchi-Bergmann 1994) with $E(B-V)$ varying from 0.1 to 2.5, in 
steps of 0.1. 

The results are presented in Table 1, which lists in columns (1) through (7) 
the galaxies' identification number in our catalog, their coordinates, the 
best-fit photometric redshift and template type, the observed magnitude in the 
NICMOS/F160W bandpass, and the observed optical and near-infrared colors.
We found that all but three (\#5256, \#6548, and \#1927) of the original nine 
candidates remain best characterized by an E/S0 or Sab template.  Within the 
plausible range of extinctions we explored, dust reddening cannot explain the 
red SEDs of galaxies \#12182, \#9024, \#9151, \#6140, \#1223, and \#12183.   We
conclude that these six galaxies are genuinely evolved populations at $z>2.8$.
Galaxy \#5256 was found to be better described as a dusty starburst at $z=1.88$
with $E(B-V)=1.3$; galaxy \#6548 was identified as a dusty starburst at $z=
1.58$ with $E(B-V)=1.3$; and galaxy \#1927 was identified as a dusty starburst 
at $z=3.43$ with $E(B-V)=1.6$.   The lower redshifts for \#5256 and \#6548 are 
further supported by the presence of significant flux in the ACS/F435W 
bandpass, which is inconsistent with Lyman limit absorption at $z>3.7$.  

A summary of the redshift likelihood analysis of the six galaxies best-fitted 
by early-type galaxy templates is presented in Figure 2 together with their ACS
and NICMOS images.  Using the photometric redshifts derived from the expanded 
template set, we now proceed to examine the physical properties of these six 
galaxies in detail.  We discuss in turn the age estimate of the stellar 
population, the accumulated stellar mass, the star formation rate, and the 
morphology.

{\em Age}---In principle, the age of a galaxy $t_0$ can be determined through a
$\chi^2$ analysis.  The probability that the observed SED is consistent
with a spectral synthesis model $k$ at a given age $t$ is defined as 
$p(k,t)=\prod_{i=1}(\sqrt{2\pi}\sigma_i)^{-1}\exp\left[-(f_i-F_i(k))^2 /
2\sigma_i^2\right]$,
where $i$ is the number of bandpasses available, $f_i$ and $\sigma_i$ are the 
observed flux and flux error in bandpass $i$, and $F_i(k)$ is the corresponding
predicted flux from model $k$.  The model spectra may span a range of star
formation history and metallicity.  Although the best-fit age is subject to 
known degeneracies between different physical properties---most notably the 
age-metallicity degeneracy---the range of possible ages is bounded.  The 
minimum age consistent with a given SED may be determined, for example, from 
the maximum plausible metallicity.  Regardless of the detailed star-formation 
history, the minimum in the luminosity-weighted mean age is robust if a 
plausible upper limit to the metallicity can be identified.  At these early 
epochs, solar metallicity would appear to be a reasonable upper limit for the 
average stellar population observed from the integrated light of individual 
galaxies.  


We performed a minimum $\chi^2$ analysis to estimate the ages of the red 
galaxies, imposing the constraint that the metallicity be less than or equal to
solar and considering a range of ages from 1 Myr through the age of the 
Universe at the redshifts of the galaxies.  The estimated minimum ages (which
corresponds to the maximum plausible metallicity) are presented in column (8) 
of Table 1.  The best-fit model is identified in column (9).  Our analysis 
suggests that these galaxies are genuinely old with a likely formation epoch 
well beyond $z_f=10$.  It is, however, interesting to find that only two of 
these galaxies can be described by a single burst model, indicating that some 
star formation is still in progress (as shown more quantitatively below).  We 
note that the results of our analysis are only mildly sensitive to the 
uncertainties in photometric redshifts.  Lowering the redshift by $\sim 0.3$ 
(typical of the observed scatter of photometric redshifts at $z \approx 3$ due 
to template mismatch uncertainty) reduces the estimated age by only 0.3 Gyr.

{\em Stellar mass}---We estimated the total stellar mass of the six 
near-infrared selected galaxies using the observed flux in the NICMOS/F160W
filter and the mass-to-light ratio $M/L_B$ of the best-fit spectral synthesis 
model at the estimated age.  The NICMOS/F160W band corresponds roughly to the 
rest-frame $B$ band at $z\sim 3$, and the adopted values of $M/L_B$ ranged from
$M/L_B=0.7$ for model (3) in Table 1 to $M/L_B=2$ for model (2).  The resulting
stellar masses are presented in column (10) of Table 1.  The estimated masses 
range from $3.5\times 10^{9}\,h^{-2}\,{\rm M_\odot}$ to more than 
$10^{11}\,h^{-2}\,{\rm M_\odot}$ by $z=2.9$, comparable to the stellar mass 
observed in nearby, bright elliptical galaxies.  For comparison, the 
characteristic stellar mass measured at $z=0$ is $M_*=7.07\times 
10^{10}\,h^{-2}\,{\rm M_\odot}$ averaged over all galaxy types (Cole \etal\ 
2001).  We note again that lowering the redshift of each galaxy by the typical 
uncertainty in the photometric redshift ($\sim 0.3$) reduces the estimated 
stellar mass by no more than 25\%.

{\em Star formation rate}---From the age-fitting analysis, we found that only 
two of these red galaxies can be explained by a single burst at higher 
redshift.  The moderate fluxes observed in the ACS bandpasses (which correspond
to rest-frame near-UV bandpasses at $z>3$) indicate that at least some star 
formation is proceeding in these galaxies.  We derived the intrinsic SFR using 
the UV SFR estimator at 1500 \AA\ published by Madau, Pozzetti, \& Dickinson 
(1998).  The results are presented in column (11) of Table 1.  The SFR of the 
evolved galaxies ranges from 0.04 to 5 ${\rm M_\odot} {\rm yr}^{-1}$, while the
SFR of the dusty starburst galaxies fall in the range $2-30\,{\rm M_\odot} {\rm
yr}^{-1}$ after corrected for dust extinction.  The derived rates distinguish
these evolved galaxies from the massive, star-forming population at $z\sim 2$
(e.g.\ Daddi \etal\  2004; Savaglio \etal\ 2004), but show that they continue 
to form stars at rates comparable to those observed in present-day spiral 
galaxies (e.g.\ Kennicutt 1998).

{\em Morphology}---Galaxy \#1223 in Figure 2 appears quite compact in the 
rest-frame $B$ band (observed in the NICMOS/F160W band) but more extended in 
the rest-frame ultraviolet (observed in the ACS images).   As a first step 
toward quantifying the morphologies of these galaxies, we calculated central 
concentration indices following the procedure outlined in Abraham \etal\ 
(1994).  Figure 3 compares the concentration indices measured for the six 
evolved galaxies listed in Table 1 to those of all galaxies detected on the
NICMOS/F160W frame.  In the rest-frame $B$ band, galaxy \#1223 is more 
concentrated than any other galaxy at comparable redshift.  Although the 
absolute values of the concentration indices vary with detection threshold and 
redshift (due to cosmological surface-brightness dimming), the ordering of the
concentrations at constant redshift does not.  Both the large relative 
concentration and the visual appearance of galaxy \#1223 suggest that much of 
its stellar mass is confined to a compact component.

In the rest-frame UV, however, galaxy \#1223 is more diffuse.  A closer 
inspection of this object shows that the lower concentration in F775W results 
from an asymmetric, nearly linear distribution of UV-bright sources nearby.  
The large scatter in F775W concentration at fixed F160W concentration for the 
NICMOS-detected sample as a whole suggests that such differences between UV and
optical morphology are common.  The concentration indices of the other five 
evolved galaxies identified in this work appear to be typical of the 
NICMOS-detected sample in both filters.


\section{DISCUSSION}

We have analyzed the HUDF images and identified six red galaxies with $AB_{\rm
F160W} \le 26.5$ at $z>2.8$, three of which have been studied previously by 
Caputi \etal\ (2004) based on shallower-depth images.  The observed SEDs are 
inconsistent with reddened, moderate-redshift galaxies and are 
best-characterized by evolved galaxy templates older than 1.6 Gyr with moderate
ongoing star formation.  Further comparison between the observed SEDs and the 
best-fit stellar synthesis models shows that four of these galaxies have 
accumulated stellar masses comparable to those seen in present-day $L_*$ 
galaxies.  The existence of these massive, evolved galaxies at $z>2.8$ is an 
important constraint for models based on hierarchical galaxy formation.  

We first examine whether active galactic nuclei (AGN) contribute to some of
these massive evolved galaxies (e.g.\ Cowie \etal\ 2001).  Comparison of
the coordinates of the six evolved galaxies with those in the 1 Ms 
catalog published by Giacconi \etal\ (2002) shows that one of the six 
evolved galaxies, \#1223, has detections both in the $0.5-2$ keV and $2-10$ keV
bands to within the $1.5''$ error radius\footnote{It has come to our attention 
after the paper was submitted to the {\em Journals} that this source had been
observed spectroscopically by Szokoly \etal\ (2004) and identified as type 2
QSO at $z=3.064$ based on the detections of prominent Ly$\alpha$ and C\,IV 
emission features (XID\#27).}.  The X-ray luminosity of this galaxy over the 
rest-frame $2-8$ keV window is estimated to be $3.6 \pm 0.4 \times 10^{43}$ erg
s$^{-1}$ at $z=3.46$.  The luminous X-ray flux together with its concentrated 
morphology indicates that this galaxy has an active nuclear region.  We cannot 
assess whether the nuclear region is heavily obscured by dust due to the lack 
of soft X-ray data in the rest frame.  Given that both dusty and QSO templates 
are considered a mismatch to the observed SED, we argue that the active nucleus
is unlikely to contribute a large fraction of the observed optical and 
near-infrared fluxes.  But because it is difficult to assess the AGN 
contribution to the observed broad-band fluxes, we exclude this source from the
subsequent analysis to calculate the stellar mass density.
 
Next, we estimate the total, volume-averaged stellar-mass density contained in 
evolved galaxies by $z=2.5$.  Given the total stellar mass determined for each
evolved galaxy, we first calculate the maximum accessible comoving volume at 
$z> 2.5$ of the images based on the expected NICMOS/F160W brightness derived 
from the best-fit spectral synthesis model and the corresponding $M/L_B$.  
Individual contributions from the five evolved galaxies (excluding the possible
AGN contamination from \#1223) are summed together to yield a total, comoving 
stellar-mass density.  We estimate that the total stellar mass density confined
in evolved galaxies by $z=2.5$ is at least $\rho_{\rm star}=0.7-1.2\times 
10^{7}\,h\,{\rm M_\odot}\,{\rm Mpc}^{-3}$.  The primary source of uncertainty 
in $\rho_{\rm star}$ is the uncertainty in $M/L_B$ for different star formation
histories and metallicities.

Our estimate of the stellar mass density at $z>2.5$ accounts for galaxies of 
total stellar mass greater than $M_0 = 10^{9}\,h^{-2}\,{\rm M_\odot}$ if we 
adopt $M/L_B=2$.  Below this mass threshold, the galaxies will have $AB_{\rm 
F160W}> 27.8$ and will therefore fall below the detection limit of the 
HUDF/NICMOS images.  This estimate is $\approx 16-28$\% of the total stellar 
mass density reported by Dickinson \etal\ (2003) and Rudnick \etal\ (2003) at 
$z\sim 3$, suggesting that massive, evolved galaxies account for a substantial 
fraction of the stellar mass density even at $z>2.5$.

van Dokkum \etal\ (2003) have reported a surface density of $0.9 \pm 0.2$ 
arcmin$^{-2}$ for red galaxies of $K < 21$ and $J-K>2.3$ at $z\ge 2$, based on 
observations of $2.3\times 2.3\,{\rm arcmin}^2$ sky area (c.f. Saracco \etal\ 
2004).  Excluding galaxy \#1223, we find none of the evolved galaxies at $z >
2.5$ are brighter than their thresold.  Given the small field sizes and the 
strong clustering measured for evolved galaxies at lower redshift, this 
difference could simply be the result of field-to-field fluctuations.  It is 
interesting to note that sub-mm observations have revealed a number of massive 
galaxies at $z\sim 2.5$ in the process of rapidly converting baryons into stars
(e.g.\ Genzel \etal\ 2003).  It appears that observations at different 
wavelengths offer snapshots of these galaxies at different stages of their 
lifetime as the universe ages, and our sample may be among the ones that 
finished the major episode of star formation the earliest.

In summary, we have identified five evolved galaxies and one active galaxy at 
$z>2.8$ in the central 5.76 arcmin$^2$ HUDF/NICMOS region based on a pronounced
broad-band discontinuity between the NICMOS F110W and F160W bandpasses.  These 
galaxies have red observed optical and near-infrared colors ($AB_{\rm F775W}-
AB_{\rm F160W} \ge 2$) that cannot be explained entirely by dust reddening 
(indicating advanced age), and their luminosities at rest-frame optical 
wavelengths indicate a substantial stellar mass.  Our analysis suggests that 
these galaxies have stellar masses comparable to the present-day $M_*$ and are 
at least 1.6 Gyr old at $z>2.8$, when the universe was merely 2 Gyr old.  
Because of their modest star-formation rates, all five evolved galaxies have 
$AB_{\rm F775W}> 26$ and thus would not have been included in samples selected 
by rest-frame UV features (e.g.\ Shapley \etal\ 2001).  The presence of 
massive, evolved galaxies at early times suggests that early star formation may
have been particularly efficient in massive halos (e.g.\ Stockton, Canalizo,
\& Maihara 2004; Glazebrook \etal\ 2004).

\acknowledgments

We are grateful to the Hubble Ultra Deep Field working group at the Space 
Telescope Science Institute led by Dr.\ Steve Beckwith and the HUDF NICMOS 
treasury project team led by Prof.\ Rodger Thompson for making the calibrated 
and carefully drizzle-stacked images available to the public.  We thank Paul 
Schechter and Lynn Matthews for helpful comments on an earlier draft; Andrew 
Baker and Karl Glazebrook for interesting discussions; and the referee Dr. 
Bahram Mobasher for helpful comments.  H.-W.C.\ acknowledges support by NASA 
through a Hubble Fellowship grant HF-01147.01A from the Space Telescope Science
Institute, which is operated by the Association of Universities for Research in
Astronomy, Incorporated, under NASA contract NAS5-26555.

\clearpage

\begin{deluxetable}{p{0.5in}ccrrrcccrr}
\tabletypesize{\scriptsize}
\rotate
\tablecaption{Properties of red galaxies identified in the HUDF} 
\tablewidth{0pt}
\tablehead{\colhead{} & \colhead{RA} & \colhead{Dec} &  \colhead{} &  
\colhead{} & \colhead{} & \colhead{$AB$} & 
\colhead{$t_{\rm min}$} & \colhead{} & \colhead{Stellar Mass} &
\colhead{SFR} \\
\colhead{ID} & \colhead{(J2000)} & 
\colhead{(J2000)} &  \colhead{$z_{\rm phot}$} &  \colhead{SED} & 
\colhead{$AB_{\rm F160W}$} & \colhead{$I-H$\tablenotemark{a}} & 
\colhead{(Gyr)} & \colhead{Model\tablenotemark{b}} & 
\colhead{($h^{-2}\,{\rm M_\odot}$)} & 
\colhead{($h^{-2}\,{\rm M_\odot}{\rm yr}^{-1}$)} \\
\colhead{(1)} & \colhead{(2)} & \colhead{(3)} & \colhead{(4)} & \colhead{(5)} &
\colhead{(6)} & \colhead{(7)} & \colhead{(8)} & \colhead{(9)} & \colhead{(10)}&
\colhead{(11)}}
\startdata
12182  & 03:32:42.88 & $-$27:48:09.5 & 2.91 & E/S0 & $24.752\pm 0.015$ & 4.7 & 2.2 & (2) & $3.1\times 10^{10}$ & $0.04\pm 0.01$ \nl
09024  & 03:32:35.08 & $-$27:46:47.5 & 3.11 &  Sab & $23.852\pm 0.008$ & 2.4 & 1.7 & (3) & $3.2\times 10^{10}$ & $1.01\pm 0.02$ \nl
09151  & 03:32:43.36 & $-$27:46:47.2 & 3.29 &  Sab & $26.529\pm 0.050$ & 2.1 & 1.7 & (3) & $3.5\times 10^{9}$ & $0.12\pm 0.01$ \nl
06140  & 03:32:43.48 & $-$27:47:27.3 & 3.33 &  Sab & $25.369\pm 0.013$ & 2.1 & 1.6 & (3) & $9.9\times 10^{9}$ & $0.41\pm 0.01$ \nl
01223  & 03:32:39.67 & $-$27:48:50.6 & 3.46 &  Sab & $22.593\pm 0.003$ & 2.0 & 1.7 & (3) & $1.8\times 10^{11}$ & $5.23\pm 0.03$ \nl
12183  & 03:32:38.74 & $-$27:48:39.9 & 4.25 & E/S0 & $25.313\pm 0.024$ & 3.8 & 1.4 & (2) & $9.1\times 10^{10}$ & $0.19\pm 0.04$ \nl
\tableline 
05256  & 03:32:42.74 & $-$27:47:33.9 & 1.88 & $E(B-V)=1.3$ & $23.141\pm 0.004$ & 2.4 & ... & ... & ... & $6.23\pm 0.27$ \nl
06548  & 03:32:34.64 & $-$27:47:20.9 & 1.58 & $E(B-V)=1.3$ & $23.815\pm 0.007$ & 2.2 & ... & ... & ... & $1.92\pm 0.14$ \nl
01927  & 03:32:39.17 & $-$27:48:32.4 & 3.43 & $E(B-V)=1.6$ & $23.907\pm 0.004$ & 2.6 & ... & ... & ... & $28.2\pm 0.5$ 
\enddata
\tablenotetext{a}{The observed optical and near-infrared color measured in the
ACS/F775W and NICMOS/F160W bandpasses.}
\tablenotetext{b}{The spectral synthesis models were generated using the
Bruzual \& Charlot (2003) stellar synthesis code with a Salpeter initial mass
function and the following recipes for SFR history and metal content: (1) 
single burst with $1/5$ solar metallicity; (2) single burst with solar 
metallicity; (3) $\tau=0.3$ Gyr SFR with $1/5$ solar metallicity; and (4) 
$\tau=0.3$ Gyr SFR with solar metallicity.}
\end{deluxetable}

\newpage

\begin{figure}
\epsscale{0.75}
\plotone{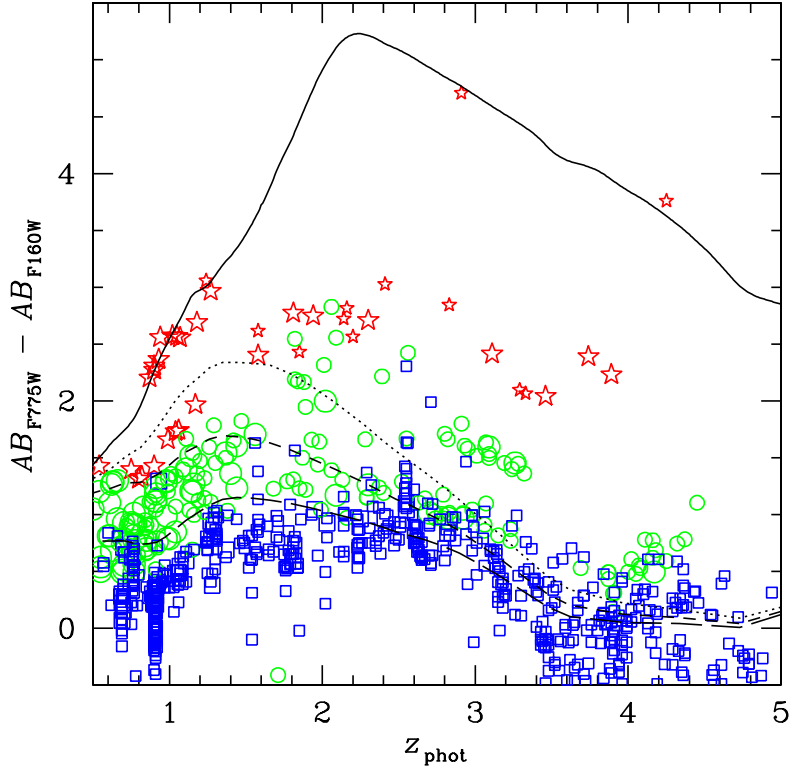}
\caption{The observed ACS/F775W $-$ NICMOS/F160W color versus photometric 
redshift measurements for 1833 galaxies identified in the HUDF/NICMOS images.
Different symbols represent different best-fit template types, from early-type 
templates (stars), through intermediate-to-late type (open circles), and 
through starburst templates (squares).  Larger symbols represent galaxies with
an observed F160W flux brighter than $AB=23.9$ ($1\,\mu$Jy), and smaller ones
are for the fainter galaxies.  The curves show the expected optical and near-ir
color evolution with redshift for a non-evolving elliptical/S0 template (solid 
curve), an exponentially declining SFR model with a formation redshift of $z_f=
30$ and e-folding times of 1 Gyr (dotted curve) and 2 Gyr (short dashed curve),
and constant SFR model (long dashed curve).  The apparent segregation of points
at $z_{\rm phot}\approx 0.9$ and 2.6 is due to the lack of $U$-band photometry
for the HUDF galaxies that leads to confusions between the 4000-\AA\ 
discontinuity from low-redshift galaxies and the Ly$\alpha$ discontinuity from
high-redshift galaxies.  These galaxies have relatively blue SEDs and thus do 
not contribute to the evolved galaxy population at all redshifts.}
\end{figure}

\newpage

\begin{figure}
\epsscale{0.9}
\plotone{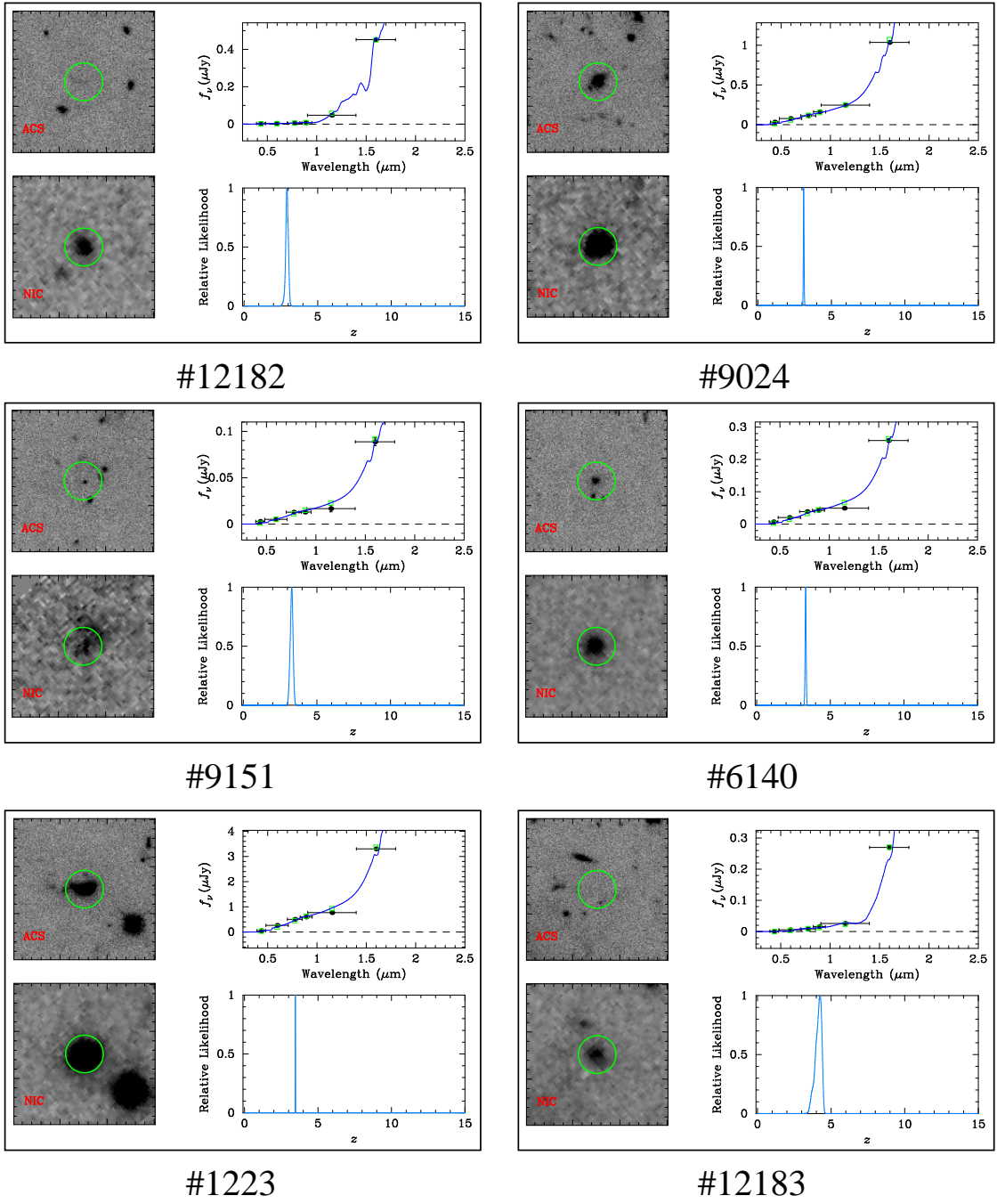} 
\caption{Summary of our photometric redshift analysis for the six evolved
massive galaxies identified at $z>2.8$ in the Hubble Ultra Deep Field.  In
each panel, the combined ACS and NICMOS images are presented on the left.  The
image extent is $4.5''$ on a side.  The observed spectral energy distribution 
established based on optical and near-ir broad-band photometric measurements is
presented in the upper-right corner of each panel, together with the best-fit 
template (solid curve) and model fluxes (open squares).  The redshift 
likelihood function for each galaxy is presented in the lower-right corner of 
each panel, indicating the most likely photometric redshift according to the 
likelihood analysis.  We see in every case that the large flux decrement 
between the NICMOS F110W and F160W is identified as the 4000-\AA\ break in the 
rest frame.}
\end{figure} 

\newpage

\begin{figure}
\epsscale{0.9}
\plotone{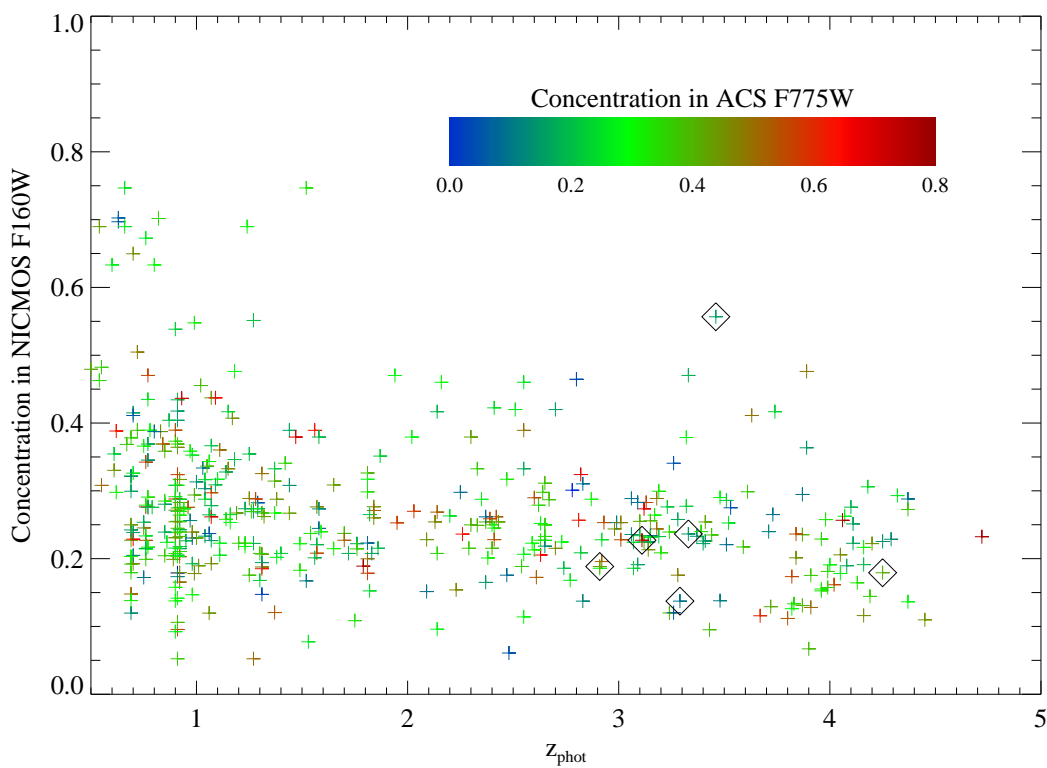} 
\caption{Central concentration indices measured for the NICMOS detected
objects in the F160W band versus photometric redshift.  The concentration 
measured for the rest-frame UV light (in the ACS/F775W band) is shown in colors
as indicated by the color bar in the upper-right corner.  Comparison of the six
evolved galaxies (indicated by the diamonds) with the field sample at similar 
redshift shows that galaxy \#1223 is the most concentrated of the sample in the
rest-frame optical but is moderately diffuse in the rest-frame UV.}
\end{figure}

\end{document}